\documentstyle[prl,aps]{revtex}
\begin{document}
\draft
\twocolumn[\hsize\textwidth\columnwidth\hsize\csname
@twocolumnfalse\endcsname

\title{Coiling of Cylindrical Membrane Stacks   
with Anchored Polymers}
\author{Vidar Frette,$^{1,}$\cite{vidaradd} Ilan Tsafrir,$^1$
Marie-Alice
Guedeau-Boudeville,$^2$
Ludovic Jullien,$^3$\\ Daniel Kandel$^1$ and Joel Stavans$^1$}
\address{\vspace*{3 mm} $^1$Department of Physics of Complex Systems, 
The Weizmann Institute of Science,\\
Rehovot 76~100, Israel. \\
      \vspace*{3 mm}
$^2$Laboratoire de Physique de la Mati\'{e}re Condens\'{e}e, URA 792,
Coll\`{e}ge de France,\\
11 Place Marcelin Berthelot,
F-75231 Paris CEDEX 05, France. \\
      \vspace*{3 mm}
$^3$D\'{e}partement de Chimie de l'\'{E}cole Normale Sup\'{e}rieure, URA
1679,
24, rue Lhomond,\\
F-75231
Paris CEDEX 05, France.  \\
}
\maketitle
\vspace*{2 mm}

\begin{abstract}
We study experimentally a coiling instability of cylindrical
multilamellar stacks
of phospholipid membranes, induced by 
polymers with hydrophobic anchors grafted along their hydrophilic
backbone. 
We interpret our experimental results in terms of a model in which
local membrane curvature and polymer concentration are coupled.
The model predicts the occurrence of maximally tight coils above a
threshold polymer concentration. Indeed, only maximally tight coils are
observed experimentally. Our system is unique in that coils form
in the absence of twist and adhesion.
\end{abstract}
\vspace{0.3cm}
{\hspace*{2.cm}}PACS numbers: 87.16.Dg, 68.10.-m
\vspace{0.3cm}
]
\input epsf
\newpage

The coil motif is ubiquitous in a wide range of natural contexts. 
One-dimensional filaments
of mutant bacteria \cite{Mendelson:78}, supercoiled DNA
molecules \cite{Cozzarelli}, and tendrils of climbing plants
\cite{Goriely:98}
all exhibit a writhing instability as a result of forcing or interaction
with
an external agent. Such
systems are dominated by elastic properties and 
the appearance of coils is a result of the relief
of twist.
In this paper we show that coiling can also be effected in
cylindrical multilamellar tubes of phospholipid bilayers, by anchoring
hydrophilic polymers with
hydrophobic side groups grafted along the backbone. 
This system is unique in that, in contrast with the above examples,
fluid membranes cannot support any twist. Yet coils are formed in the
system, and are stable for a very
long time.

Our system is representative of 
a wide class of systems of membranes with embedded inclusions, such as
biological cells with membrane-associated proteins.
Other examples include erythrocyte ghosts incorporating amphipathic
drugs
\cite{sheetz:74} and liposomes with covalently attached polymers
used for drug delivery \cite{sunamoto:84,lasic:94,Yanga}
(for additional examples see \cite{Ringsdorf:88}). Our
experiments give us a
unique opportunity to study a wide variety of phenomena induced by
such inclusions in a relatively simple and controlled environment.

The cylindrical multilamellar tubes,
called {\it myelin figures}, consist of a large number
of bilayers reaching almost to the core \cite{Lehmann:13,Sakurai:89}. 
Adjacent bilayers are
separated by thin hydration layers.
Coiling of myelin figures has already been observed in
phospholipid binary mixtures \cite{Lin} when the constituents undergo a
phase separation process triggered by the addition of Ca$^{++}$.
Coiled myelin figures of egg-yolk
phosphatidylcholine (egg-pc) have also been reported 
\cite{Sakurai:85,Sakurai:85a}.
In both cases \cite{Lin,Sakurai:85,Sakurai:85a} coiling was attributed
to surface adhesion. The novelty of our work is that coiling occurs in
the absence of both adhesion and twist. 
Our experiments clearly show that
surface adhesion is negligible in our system.

\begin{figure}[h]
   \epsfxsize=75mm
   \centerline{\epsffile{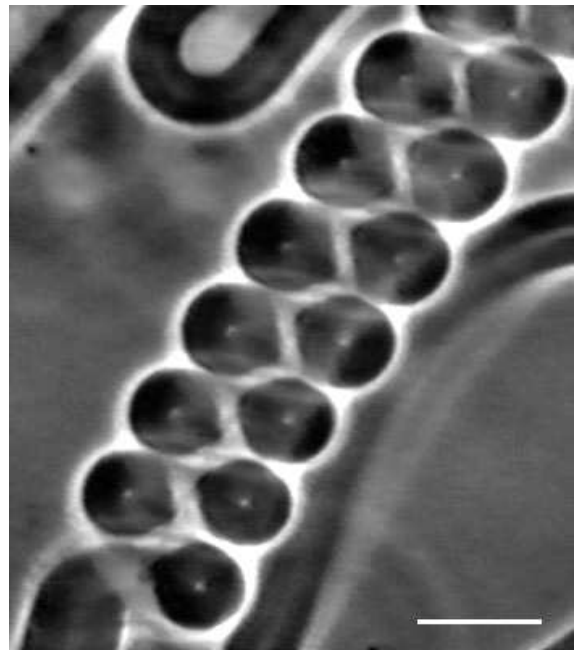}}
   \vspace{0.5cm}
\caption{A single-stranded coil. 
This structure formed one hour after hydration in a solution 
containing 
0.02 mg/ml polymer, at a rate of 0.15  
pitch/min.  
Phase contrast images a slice of the coil, showing bilayers filling up
almost the whole tube, though a narrow water-filled core
can
still be observed along the
tube axis. The bar represents 10$\,
\mu$m.
}
\end{figure}

We believe that in our system coiling results from a coupling between
the polymer concentration and local membrane curvature, induced by 
anchoring of the polymer in the membrane. We show below that
a high enough polymer density together with the constraints imposed by
the cylindrical geometry of the tube destabilizes the straight tube.
Coiling as a manifestation of this instability is proposed here for the
first time.

The coupling between membrane curvature and polymers has been considered
both theoretically 
\cite{Brooksa,Lipowsky:95,Lipowsky:95a,Lipowsky:95b} 
and experimentally
\cite{Yanga,Ringsdorf:88,Simon:95,Yang,degennes:90,diederich:99}. The
theoretical
studies emphasize the polymer backbone and
its effect on the elastic properties of the membrane. They do
not consider the mobility of anchors embedded in the bilayers. We
believe this effect is important in our system, and attempt to capture
it in the model presented below.

\begin{figure}[h]
   \epsfxsize=80mm
   \centerline{\epsffile{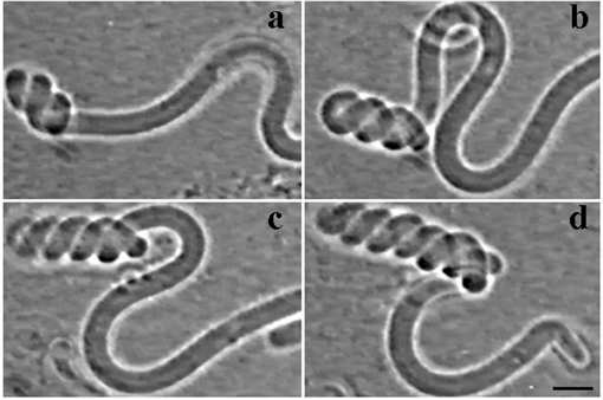}}
   \vspace{0.5cm}
\caption{Formation of a double helix. After bending, the tip slides
along 
the tube,
coiling at the same time. The four frames were taken at times (a) 0, (b)
20, (c) 30 and (d) 40 minutes after the double helix was first seen.
The bar represents 20$\, \mu$m.}
\end{figure}

In our experiments, tubular membrane stacks were made of
stearoyl-oleoyl-phosphatidylcholine
  (SOPC), with \(C_{18}\) alkyl chains. The polymer we used is
  hydrophilic
  dextran (MW 162,000 g/mol) functionalized both with \(C_{16}\) alkyl
  chains
  and dodecanoic NBD chains as fluorescent markers. The hydrophobic
  anchors,
  distributed statistically along the backbone (about 1 alkyl chains for
  25
  glucose units) are \(C_{16}\) long. On average there are 4 persistence
  lengths
  between consecutive anchors. 
  Therefore, the extension of each polymer molecule on the
  two-dimensional membrane is much larger than its extension into the
  third dimension.
  Samples were prepared by drying a 0.5-1.0\,$\mu$l droplet of SOPC
  dissolved
  in a 4:1 chloroform-methanol solution (7.35\,mg/ml) on a glass slide.
  The
  sample was then closed, and hydration was
  effected by
  injecting a polymer solution of known concentration, $c_p$, into the
  cell. The development of
  myelin
  structures and their coiling were followed using phase contrast
  microscopy
  and recorded on video. Our experiments were conducted at room
  temperature,
  well above the solid-liquid transition for SOPC.

For small values of  
\(c_{p}\) we observe myelin
figures, which display a clear tendency to straighten over lengths many
times
larger than their diameter.
As \(c_{p}\) is increased, myelin figures become
more floppy and curved.
For large enough values of \(c_{p}\), a writhing instability sets in and
tubes bend, forming irregular structures, single coils (Fig.\
1) and
double helices (Fig.\ 2).
We emphasize that all the coiled structures we observe are 
maximally tight as they form and do not tighten up gradually
(see Fig.\ 2); no loose coils have been found
(unlike Sakurai et
al. \cite{Sakurai:85,Sakurai:85a}).  
In quantitative terms, 
this means that the curvature of the tube central line, $C$, is 
$C\approx\frac{1}{r_0}$, where $r_0$ is the radius of the tube.
Fig.\ 2 also demonstrates that there is no adhesion between membrane
surfaces in our system; tube segments that are in contact in Fig.\ 2b
are separated in Fig.\ 2c.

We stress that while the polymer
concentration in solution, $c_p$, is known,
we do not control the surface concentration
on the bilayers. The slow evolution of some of the structures we observe
is consistent with a possible variation of this concentration over time. 
As a first
step towards a theoretical understanding of this system, we neglect this
slow evolution.

We now make the following assumptions: ({\it i})  anchors penetrate the
membrane to a depth of about
half a bilayer. This is because the anchor length is comparable to that
of a
lipid,
and the large hydrophilic backbone to which anchors are
attached cannot penetrate a
bilayer. We assume that these anchors and the polymer backbone
induce a local spontaneous curvature, $H_0$
\cite{sheetz:74,Ringsdorf:88,Helfrich:94}.   ({\it ii}) Polymer
molecules
are present everywhere in
the system, including
inner regions of the myelin figures. We have carried out fluorescence
microscopy experiments and found 
that the polymer is present between layers 
of the stack in significant concentrations
(additional details of the  
experiments will be published elsewhere). 
({\it iii}) Given that bilayers are in a
liquid-like state, polymer molecules can
diffuse on the membrane. Thus, the energy of the system is lowered
if they migrate to regions 
where the mean surface curvature  $H$   is
closer to $H_0$. ({\it iv}) Bilayers in a myelin figure maintain a
constant area and
are strongly constrained by the geometry of the
stack, which prevents them from buckling.
We therefore regard the myelin figures as flexible cylinders having a
fixed circular cross section everywhere along their axis. These
assumptions are strongly supported by our experimental observations.
Note that the central line of the tube can bend. However, as a
consequence of the above assumptions, its length remains fixed.

Based on these assumptions, we present a simple model
which captures the essential physical features of the experimental
system, and
accounts for many of the experimental observations. We
represent each bilayer as two square lattices (in the spirit of
lattice-gas models), corresponding to the outer and inner monolayers.
We associate two degrees of freedom with each lattice site, which
corresponds to a membrane patch of area
$a^2$. The first is the local mean surface
curvature, $H$. The second  is a
binary occupation variable, which takes the values one
or zero when the site is occupied or unoccupied respectively
by an anchored molecule. This molecule induces a local spontaneous
curvature, $H_0$. 
The area of a site,
$a^2$, and the value of $H_0$ depend on the specific mechanism
responsible for the spontaneous curvature. Thus, if $H_0$ is induced by
individual anchors, the area of a site is microscopic ($a^2\sim 60{\mbox
\AA}^2$). If, on the other hand, the spontaneous curvature is induced by
the polymer backbone, the area of a site is mesoscopic ($a^2\sim 6\cdot
10^5{\mbox \AA}^2$).

By convention, the curvatures of
the inner and outer layers have opposite signs at the same position.
Within the model, the energy of the system is a sum of the curvature
energies of the individual area
patches: $2\kappa H^2$ for a
vacant site, and $2\kappa^{'}(H-H_0)^2$ for an occupied site.
$\kappa$ and $\kappa'$ are the local 
bending rigidities of a single layer without and with an anchored
molecule, 
respectively. We suppose $\kappa^{'} > \kappa$; this is consistent with 
models of composite membranes
\cite{Lipowsky:95a,Lipowsky:95b,Helfrich:94}
(although the systems these models describe are different from ours).

In order to find the equilibrium state of a tube, we have to calculate
its free energy. 
This free energy depends on the curvature of its
central line, $C$, and on $\rho$, the average of the occupation
variable ($\rho a^2$ is the average density of the anchored molecules). 
If the spontaneous curvature, $H_0$, is large enough,
the free energy of a bent tube is lower than that of a straight one. To
show this we evaluate separately the energy and the
entropy of the system.

Consider one cylindrical bilayer of length $l$
and circular cross section of radius $r$, with the same average 
occupancy, $\rho$, on both sides. Let us calculate the
energy cost of bending the bilayer into a portion of a coil with
central line curvature $C$ in two steps. First, the
energy of a bent cylindrical bilayer with a {\it
homogeneous}
distribution of anchored molecules is $E_{hom}=E_{hom}^{out} +
E_{hom}^{in}$, where
$E_{hom}^{out,in}$ are the energies of the outer and inner monolayers.
According to our model
\begin{eqnarray}
E_{hom}^{out,in}(C)&=&2\rho\kappa'\int
dA\left[H(C)-H_0\right]^2\nonumber\\
&+&
2(1-\rho)\kappa\int
dA\left[H(C)\right]^2~,
\label{ehom1}
\end{eqnarray}
where $H(C)$ is the local membrane curvature, and is known for
a cylindrical geometry. The expressions for $E_{hom}^{out}$ and
$E_{hom}^{in}$ are different
since the curvatures of the inner and outer monolayers have opposite
signs. Note that when $C\not =0$,
$H(C)$ varies around the bent cylinder. For our geometry the total mean
curvature $\int dA H(C)=2\pi l$ for the outer monolayer,
while for the inner one $\int dA H(C)=-2\pi l$, independent
of the central line
curvature $C$. Thus the cost of bending the cylindrical membrane,
keeping the distribution of anchored molecules homogeneous is $\Delta
E_{hom}=4[\rho\kappa'+(1-\rho)\kappa]\int
dA\left[H(C)\right]^2$, independent of the value of $H_0$. Note that the
term containing $H_0^2$ is independent of $C$ and therefore does not
contribute to the cost of bending.

Next, we take into account inhomogeneities in the distribution of
anchored molecules around the tube. Such inhomogeneities reduce the
energy if
these molecules move to
regions of membrane curvature closer to $H_0$ in both the outer
and inner monolayers. The full calculation shows that the energy gain,
$\Delta
E_{inhom}(C,H_0)$, depends linearly on the spontaneous curvature and can
become arbitrarily large for large values of $H_0$.

As for the entropy of the system, we assume that the
dominant contribution is the entropy of mixing of occupied and vacant
sites. This entropy is larger when the distribution of anchored
molecules around
the cylindrical bilayer is homogeneous, favoring a straight
tube. However, it does not depend on the spontaneous curvature.  
Therefore, if $H_0$ is large enough, the energy gain due to $\Delta
E_{inhom}(C,H_0)$ is larger than the
free energy cost coming from $\Delta E_{hom}$ and the entropy of mixing.
In this case, the tube is bent at equilibrium. It remains to be
shown that such an equilibrium state can occur for reasonable and
physical values of
the model parameters.

We have carried out the full calculation of the free energy as a
function of the central line curvature, the average occupancy and
the spontaneous curvature (details will be published
elsewhere). In the limit of a
thick tube (a realistic case), the calculation can
be done analytically, and
the free energy of the entire tube (summing over all the bilayers)
reads:
\begin{equation}
F(C,\rho)=\frac{2 l\kappa_{tube}(\rho)}{r_0}\cdot
\ln\left[\frac{2}{1+\sqrt{1-(Cr_0)^2}}\right]~,
\label{freeenergy}
\end{equation}
where $\kappa_{tube}(\rho)$ is the
effective bending rigidity of the entire tube, and $l$ is its length. 

Fig.\ 3 shows the typical dependence of
$\kappa_{tube}$ on $\rho$
for large enough values
of $H_0$ (solid line). When $\rho<\rho_-$, $\kappa_{tube}$ is positive
and decreases with $\rho$. 
In this regime the minimum of the free energy is at $C=0$. Therefore,
the tube is predicted to be straight on the average, but
with enhanced fluctuations due
to the smaller bending modulus.
Although anchored molecules increase the local bending rigidity of the
membrane ($\kappa'>\kappa$), their mobility makes it {\em easier} to
bend the tube.  
  
For $\rho_-<\rho<\rho_+$, $\kappa_{tube}$ is
negative, and the free energy decreases upon bending with its minimum at
the maximally possible central line curvature $C=1/r_0$ (in agreement
with the qualitative argument outlined above). Hence, $\rho_-$ is a
threshold occupancy above which straight
tubes are unstable and form {\em maximally tight} coiled structures. 
Above $\rho_+$ straight tubes become stable
again; however, this regime is
probably unreachable in our experiments, since too large a polymer
concentration destroys the bilayers.

A salient feature of our model is that it precludes loose coiled 
structures; i.e. tubes are either uncoiled ($\rho<\rho_-$ and
$\rho>\rho_+$) 
or {\em maximally tight} and coiled ($\rho_-<\rho<\rho_+$). This
is consistent with our experimental observations. In addition, according
to the model all the coils with $C=1/r_0$ are equally probable. One
should therefore expect to see irregular coils without a well defined
chirality as well as regular helices, all of which we indeed observe. 

\begin{figure}[h]
   \epsfxsize=80mm
   \centerline{\epsffile{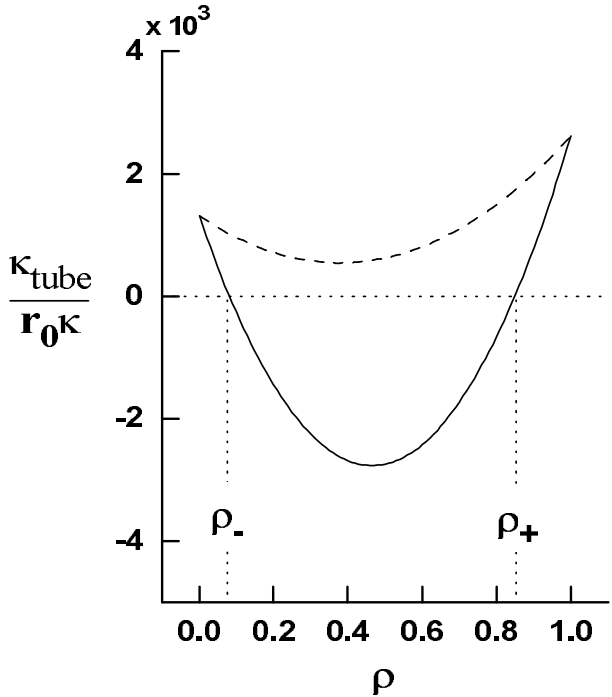}}
   \vspace{0.5cm}
\caption{The effective bending modulus of the tube, $\kappa_{tube}$, is
parabolic
in the average occupancy, $\rho$. We have used the following
values of the parameters: $\kappa=10 k_B T$, $\kappa'=2\kappa$ and
$r_0=5\mu$m. We
find that $\kappa_{tube}$ depends on $a$ and $H_0$ only through the
product $aH_0$.
The solid curve represents $\kappa_{tube}$ for 
$aH_0=0.3$. The physical meaning of this number depends on the specific
mechanism responsible for the spontaneous curvature. For example, if the
spontaneous curvature is induced by the polymer backbone, $a\sim
800{\mbox \AA}$. For the low polymer concentration in our experiments
(i.e., in the vicinity of $\rho=\rho_-$), this corresponds to a membrane
spontaneous curvature $\rho H_0\sim 0.3\mu^{-1}$. When $aH_0$ is large
enough
($aH_0>0.19$ for the values of $\kappa$
and $\kappa'$ we have used) $\kappa_{tube}<0$ between
$\rho_-$ and $\rho_+$. For smaller values of $aH_0$,
$\kappa_{tube}>0$ for all values of $\rho$. The dashed curve
corresponds to $aH_0=0.16$. 
}
\end{figure}

A somewhat similar instability of a flat membrane due to coupling
between membrane shape and local spontaneous curvature has been
discussed by Leibler \cite{Leibler} and by Safran \cite{safran:90}.
There is, however, a fundamental
difference between these models and ours. They considered an
unconstrained flat membrane, whereas our membranes are severely
constrained
by the cylindrical geometry of the tube. These geometrical constraints
play a crucial role in determining the shapes of the observed
structures,
and the final state of the system. 

We believe coiling occurs mainly due to the
spontaneous curvature induced by the anchored molecules as well as their
mobility.
Our model emphasizes these aspects, and neglects others such as
interactions between polymer molecules. We do not
expect these effects to change
the qualitative behavior of the system. 
So far, we have not been able to identify the precise mechanism by which
the polymer induces spontaneous curvature. We intend to study polymer
length effects on the coiling phenomenon in order to partially address
this issue.

We acknowledge useful exchanges with D. Bensimon, R.
Granek, R. Lipowsky, E. Moses and S. Safran. This research was supported
by
The Israel Science Foundation administered  by the Israel Academy
of Sciences and Humanities - Recanati and IDB Group Foundation.
V.F.\ gratefully acknowledges support from The Research
Council of Norway (NFR). D.K.\ is the incumbent of the Ruth Epstein Recu 
Career Development Chair.



\end{document}